\documentclass[%
 aps,
 amsmath,amssymb,
 reprint,%
superscriptaddress,
]{revtex4-1}

\usepackage{graphicx}
\usepackage{dcolumn}
\usepackage{bm}

\usepackage[utf8]{inputenc}
\usepackage[T1]{fontenc}
\usepackage{mathptmx}
\usepackage{etoolbox}

\usepackage{hyperref}

\makeatletter
\def\@email#1#2{%
 \endgroup
 \patchcmd{\titleblock@produce}
  {\frontmatter@RRAPformat}
  {\frontmatter@RRAPformat{\produce@RRAP{*#1\href{mailto:#2}{#2}}}\frontmatter@RRAPformat}
  {}{}
}%
\makeatother
\begin{document}


\title[N\'eel temperature of  RuO2]{N\'eel temperature and helical spin order of altermagnetic RuO$_2$}

\author{Markus Meinert}
 \email{markus.meinert@tu-darmstadt.de}
\affiliation{ 
New Materials Electronics Group, Department of Electrical Engineering and Information Technology, Technical University of Darmstadt, Merckstr. 25, 64283 Darmstadt, Germany
}

\date{\today}

\begin{abstract}
The magnetic groundstate of RuO$_2$ remains controversial, with experimental evidence for a nonmagnetic groundstate of ideal bulk material and indications of a magnetic state in strained thin films. Here, I investigate the N\'eel temperature of the (hypothetical) altermagnetic state of bulk RuO$_2$, stabilized via the DFT$+U$ technique, by mapping on a Heisenberg Hamiltonian. The N\'eel temperature scales monotonously with the magnetic moment up to the point where a large $+U$ term opens a band gap and turns RuO$_2$ semiconducting. The maximum N\'eel temperature obtained by this procedure is 408\,K at $U=3$\,eV, and much smaller values for smaller $U$. A reciprocal-space eigenvalue analysis reveals a helimagnetic groundstate of the spin model due to intra-sublattice antiferromagnetic coupling. This situation resembles the isostructural $\beta$-MnO$_2$, which is a prototype helimagnet. Further comparison with calculations on CrO$_2$ and altermagnetic MnF$_2$ taking $U$ as an adjustable parameter supports the validity of the spin model analysis.
\end{abstract}

\maketitle

\paragraph*{Introduction.} Ruthenium dioxide (RuO$_2$) is a rutile-structured metallic oxide that has long been classified as a Pauli paramagnet. Early semilocal density functional theory (DFT) and spectroscopic measurements supported a nonmagnetic metallic ground state hosting Dirac nodal lines and a substantial intrinsic spin Hall conductivity \cite{Sun17, Jovic18, Occhialini21}. Angle-resolved photoemission spectroscopy (ARPES) resolved these nodal lines and associated flat surface bands \cite{Jovic18}. Fermi surface measurements by quantum oscillations place additional contraints on the bulk magnetic properties of RuO$_2$ \cite{Wu2025}, thus consolidating the view of RuO$_2$ as a nonmagnetic topological metal.

This picture was challenged by neutron diffraction experiments reporting itinerant antiferromagnetism with a N\'eel temperature exceeding room temperature \cite{Berlijn17}, followed by resonant x-ray scattering that established a collinear antiferromagnetic order in both bulk crystals and thin films \cite{Zhu19}. Subsequent DFT$+U$ and many-body calculations reproduced this behavior and attributed it to a Fermi-surface Pomeranchuk instability \cite{Berlijn17, Ahn19}. These developments identified RuO$_2$ as a candidate altermagnet, characterized by symmetry-enforced, momentum-dependent spin splitting in a collinear antiferromagnet with zero net magnetization \cite{Smejkal20, Gonzalez21}. Experimental reports of anomalous Hall signals \cite{Feng2022}, spin–orbit torque generation \cite{Bose2022, Bai2022, Karube2022}, anisotropic thermal transport \cite{Zhou24}, and spectroscopic evidence for time-reversal symmetry breaking \cite{Fedchenko2024} in thin films have been interpreted within this altermagnetic framework, although other ARPES studies favor a paramagnetic description \cite{Liu2024}.

The debate has intensified with muon spin rotation and renewed neutron measurements on high-purity crystals finding no static magnetic order \cite{Hiraishi2024, Kessler2024}. First-principles calculations indicate that the energy splitting between nonmagnetic and antiferromagnetic states is small and highly sensitive to lattice parameters, doping, and correlation treatments \cite{Smolyanyuk2024}. Thin films add further variables: epitaxial strain, interfacial bonding, and structural variant selection can stabilize magnetic configurations absent in bulk. Indeed, thin-film RuO$_2$ underpins most demonstrations of altermagnetic transport signatures \cite{Zhu19, Feng2022, Bose2022, Bai2022, Karube2022}. Recent x-ray studies of single-variant RuO$_2$(101) films with controlled N\'eel-vector orientation and spin-splitting magnetoresistance highlight the impact of substrate-induced symmetry breaking \cite{He2025}.

To date, an \textit{ab initio} understanding of the N\'eel temperature of altermagnetic RuO$_2$ is missing. We attempt to fill this gap by a combination of DFT calculations, a mapping to a Heisenberg spin model, and classical Monte Carlo simulations. The altermagnetic state of RuO$_2$ is established in the calculations with the DFT$+U$ method.

\paragraph*{Calculations.} The DFT calculations were performed with the \textsc{abinit} code \cite{Amadon2008, Amadon2008a, Torrent2008, Marques2012, Gonze2020, Romero2020, Gonze2016} with the JTH PAW potentials \cite{Jollet2014} for the Perdew-Burke-Ernzerhof (PBE) \cite{PBE} functional. A $6 \times 6 \times 8$ $k$-point mesh, planewave cutoff of 20\,Ha (544.2\,eV), and tight self-consistency convergence parameters were applied. The lattice structure was taken as the experimental structure with $a=4.49$\,\AA{}, $c=3.11$\,\AA{}, and the oxygen position parameter $x=0.3058$ \cite{Kiefer2025}. The Hubbard $+U$ correction was applied with the anisotropic term $J=0$ in all calculations. Local magnetic moments are integrated within a radius of 1.17\,\AA{} around the transition metal nuclei. Maximally localized Wannier functions were generated with initial O $p$ and Ru $d$ projections with the \textsc{Wannier90} code \cite{Pizzi2020} (22 functions per spin). The Heisenberg exchange parameters $J_{ij}$ were calculated with the TB2J package \cite{He2021} up to a radius of $3a$ on a dense $k$-point mesh. The ordering vector was determined with an eigenvalue analysis in reciprocal space, and the Neel temperatures were calculated within the random-phase approximation (RPA). A multi-sublattice version of the RPA formula for a $\mathbf{q} \neq 0$ ordering vector was implemented. Details of these calculations are given in the Appendix. Classical Monte Carlo simulations of the Heisenberg Hamiltonian were performed with the \textsc{Vampire} code \cite{Evans2014, AlzateCardona2019} on $15 \times 15 \times 15$ unit cells with periodic boundary conditions, $10\,000$ steps for relaxation and $20\,000$ steps for integration. The N\'eel temperature was determined using the peak in the specific heat. Additional calculations were done for $\beta$-MnO$_2$ \cite{Bolzan1993}, MnF$_2$ \cite{Yamani2010}, and CrO$_2$ \cite{Burdett1988} with the corresponding experimental lattice constants and an otherwise identical setup.

\begin{figure}[t!]
\includegraphics[width=8.6cm]{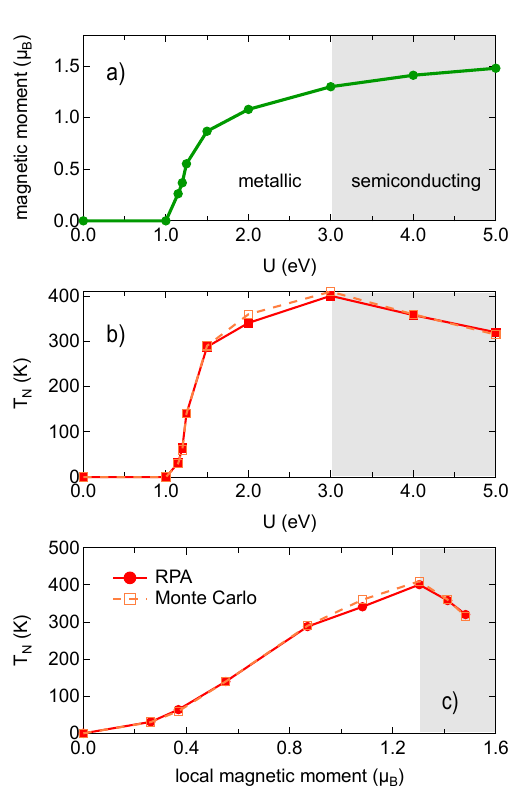}
\caption{\label{fig:RuO2_Neel_temperature_vs_U}
a): Local magnetic moment of Ru atoms as a function of the Hubbard $+U$ parameter. b) Neel temperature as a function of $U$, calculated with RPA and classical Monte Carlo simulations. c): Same as b), given as a function of the local magnetic moment.
}
\end{figure}

\begin{figure}[t!]
\includegraphics[width=8.6cm]{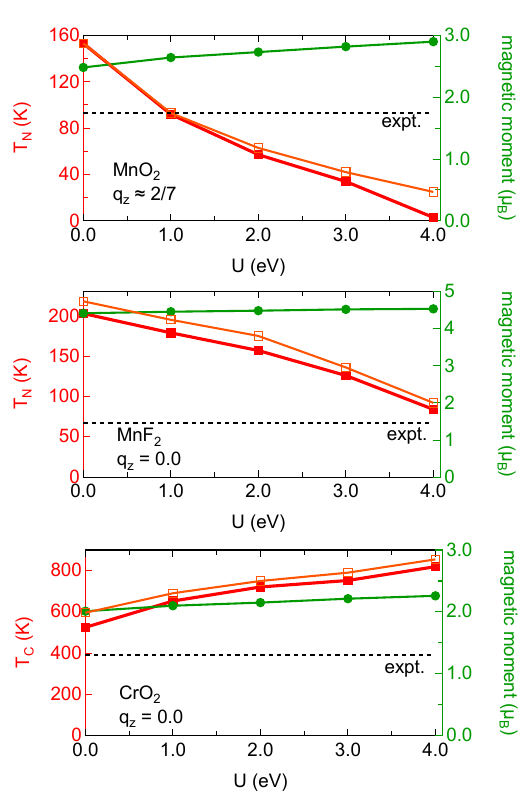}
\caption{\label{fig:Other_Neel_temperature_vs_U}
Calculations of N\'eel temperatures and magnetic moments for other rutile-structured materials: a) $\beta$-MnO$_2$, b) MnF$_2$, and c) CrO$_2$. Open symbols denote Monte Carlo results for the N\'eel temperature, full symbols are RPA results. Horizontal dashed lines indicate experimental Neel temperatures \cite{Zhou2018, Collins1974, Chamberland1977}.
}
\end{figure}

\paragraph*{Results.} In Figure \ref{fig:RuO2_Neel_temperature_vs_U} a) and b) we show the local magnetic moment and N\'eel temperature as a function of $U$. In Figure \ref{fig:RuO2_Neel_temperature_vs_U} c), we additionally plot the N\'eel temperature as a function of the local magnetic moment. We observe the increase of the local magnetic moment with $U$, with an onset just above $U=1$\,eV, in excellent agreement with previous DFT$+U$ calculations based on the PBE functional \cite{Smolyanyuk2024}. At $U=3$\,eV, the electronic structure transits from metallic to semiconducting as the on-site Coulomb repulsion opens a hybridization gap at the Fermi energy. With this band gap, RuO$_2$ resembles other rutile-structured antiferromagnets like $\beta$-MnO$_2$ and the fluorides MnF$_2$, FeF$_2$, and NiF$_2$. Because of the observation of very good metallic conductivity in RuO$_2$ \cite{Feng2022,Froehlich2002}, a large $U$ contradicts the experiment and puts a clear upper bound on the allowed value of U, i.e., $U < 3$\,eV with the PBE functional.

The agreement between the Monte Carlo calculations and our multi-sublattice RPA implementation is very good. In the metallic regime, the N\'eel temperature increases to a maximum of 408\,K at $U=3$\,eV. In the semiconducting regime, $T_\mathrm{N}$ is again reduced, although the magnetic moment further increases. For small magnetic moments, we observe a parabolic increase of $T_\mathrm{N}$. Here, the magnetization is a small perturbation of the density and the Hamiltonian of the Heisenberg model $\mathcal{H} \approx -\tilde{J}_{ij} \mathbf{m}_i\mathbf{m}_j$ (with constant $\tilde{J}_{ij}$) scales as $\mathbf{m}^2$. As the gap opens, the interaction can be better understood in terms of a tight-binding model where $J = 4t^2/\mathcal{U}$ with the hopping parameter $t$ and the on-site Coulomb repulsion $\mathcal{U}$. The Hubbard correction of the DFT+$U$ method is directly related to $\mathcal{U}$ where the double-counting renormalizes the interaction strength. By increasing $U$, the exchange term $J$ is thus suppressed and the N\'eel temperature reduces.

\begin{figure*}[t!]
\includegraphics[width=17cm]{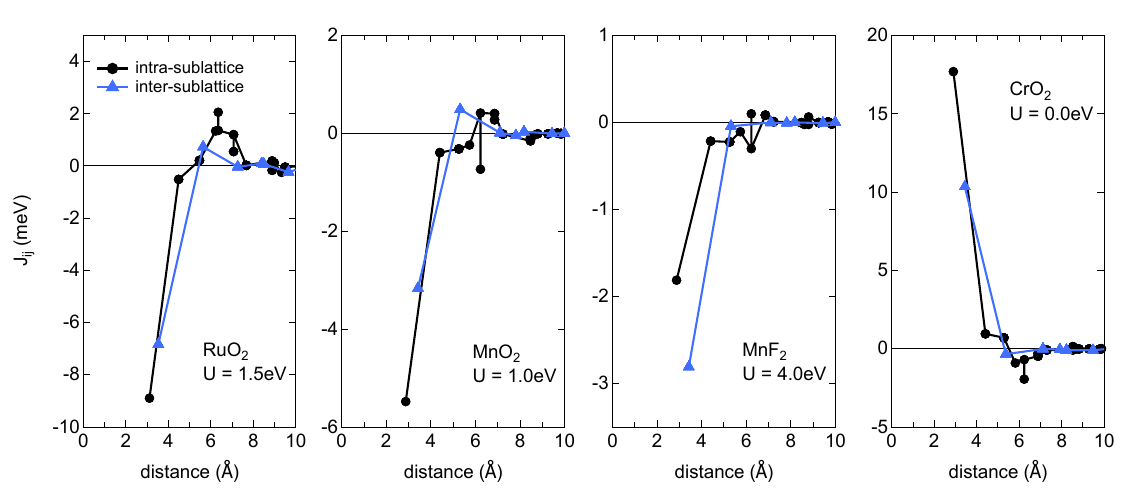}
\caption{\label{fig:Jij_data}
Real-space exchange interaction parameters $J_{ij}$ of RuO$_2$, MnO$_2$, MnF$_2$, and CrO$_2$ with $U$ values chosen such that $T_\mathrm{N}^{RPA}$ is closest to the experimental result. The interactions are resolved as intra-sublattice and inter-sublattice interactions.} 
\end{figure*}

Remarkably, the ground state of the spin Hamiltonian is not collinear for any value of $U$. In all cases, we obtain helical order with ordering vectors $\mathbf{q} = (0, 0, q_z)$ with $q_z \approx 0.178 \dots 0.32$, depending on the value of $U$ non-monotonously. Inspecting the exchange interaction parameters, we find that the origin of the helical ground state is a frustration of exchange parameters, which are antiferromagnetic both for the intra-sublattice nearest-neighbors as well as for the inter-sublattice nearest neighbors.

In the next step, we compare the magnetic order of RuO$_2$ with $\beta$-MnO$_2$, CrO$_2$, and MnF$_2$. Our calculations reproduce the observed order types (helical, collinear FM, collinear AFM) and give good agreement of calculated and experimental N\'eel (Curie) temperatures for reasonable choices of $U$. While CrO$_2$ is best described with $U=0$\,eV, $\beta$-MnO$_2$ has good agreement at $U = 1$\,eV. For MnF$_2$, a Mott insulator, a very large $U > 4$\,eV is necessary to bring the N\'eel temperature close to the experimental value. The helical order in $\beta$-MnO$_2$ has $q_z \approx 2/7 \approx 0.286$ \cite{Tompsett2012}, which is very well reproduced by the reciprocal-space eigenvalue analysis. Here the origin of the helical order is the same frustration that we observe for RuO$_2$. The dependences of $T_\mathrm{N}$ on $U$ for the metallic CrO$_2$ and the semiconducting or insulating $\beta$-MnO$_2$ and MnF$_2$ are exactly in line with our explanation for the transition behavior in RuO$_2$. The \textsc{Vampire} Monte Carlo calculations confirm the non-collinear and collinear ordering in all cases as identified by the reciprocal-space eigenvalue analysis. In all cases we also find the usually expected underestimation of the N\'eel (Curie) temperature by the RPA formula.

In Figure \ref{fig:Jij_data}, we show the exchange coupling parameters for RuO$_2$ ($U=1.5$\,eV), $\beta$-MnO$_2$ ($U=1.0$\,eV), MnF$_2$ ($U=4.0$\,eV), CrO$_2$ ($U=0.0$\,eV). The values of $U$ were chosen to give the closest fit to the experimental N\'eel (Curie) temperature, with the exception of RuO$_2$, for which we select the $U$ value that gives a local magnetic moment just below 1\,$\mu_\mathrm{B}$. Magnetic moments of similar magnitude are expected in confined mono- or bilayers of RuO$_2$ (without applying a $+U$ correction) or at the surfaces of thicker layers \cite{Ho2025, Brahimi2025}. We observe that the nearest inter- and intrasublattice exchange parameters are negative in all three rutile-structured antiferromagnets considered here, i.e. they are frustrated. In the cases of RuO$_2$ and $\beta$-MnO$_2$, the nearest intra-sublattice $J_{ij}$ (connecting equivalent Ru atoms along the crystallographic $c$ axis) is larger in magnitude than the nearest inter-sublattice interaction (connecting the corner position and the body-centered position of the tetragonal cell). In contrast, for MnF$_2$ the situation is opposite and the antiferromagnetic inter-sublattice interaction is dominant. Numerical testing on a simplified model including only the nearest intra- and intersublattice interactions (we define them as $J_1$ and $J_2$, respectively) demonstrates that a helical order is stabilized for $J_1 < J_2 < 0$, i.e. once the intra-sublattice interaction is dominant.  A smooth transition to long-wavelength spiral order is observed beyond the critical ratio, which turns into a zone-boundary staggered order along $c$ for $J_1 \gg J_2$. In the ferromagnetic case of CrO$_2$ however, both parameters turn positive and a simple collinear ferromagnetic order is the result. We note that a previous calculation of the Heisenberg parameters of RuO$_2$ showed positive nearest-neighbor intra-sublattice interaction \cite{Smejkal2023}. However, this calculation used the atomic-spheres approximation, which does not correctly reproduce the full-potential electronic structure of RuO$_2$ and fails at the correct description of the hybridization gap at large $U$.

\begin{figure}[t!]
\includegraphics[width=8.6cm]{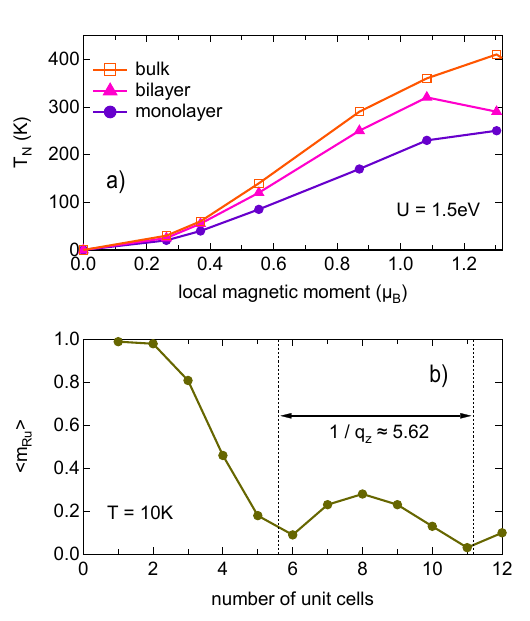}
\caption{\label{fig:bilayer_monolayer_multilayer}
a) N\'eel temperatures of RuO$_2$ obtained by Monte Carlo simulations on a monolayer (one unit cell), a bilayer (two unit cells), and bulk for comparison at $U =1.5$\,eV. b) Mean sublattice magnetization length from Monte Carlo calculations on multilayers of RuO$_2$ with $U=1.5$\,eV. Vertical dashed lines represent the expected minima positions according to the reciprocal-space eigenvalue analysis.}
\end{figure}

As theoretical modeling and experiments have shown that RuO$_2$ can host surface-stabilized \cite{Ho2025, Akashdeep2026} or confinement-induced \cite{Brahimi2025} antiferromagnetism, we use our Heisenberg Hamiltonian as a simplified model for monolayers (one unit cell) and bilayers (two unit cells). We observe that the N\'eel temperature is reduced compared to the bulk values (Fig. \ref{fig:bilayer_monolayer_multilayer} a)). Furthermore, in the confined state, the magnetic order turns collinear as antiferromagnetic intra-sublattice interactions along the $c$-axis are truncated. Thus, although the groundstate of antiferromagnetic RuO$_2$ is helical, in a geometrically confined situation this helical order is suppressed and the altermagnetic state can emerge instead. Monte Carlo simulations with open boundary conditions along $c$ demonstrate the emergence of the helical order with a period of 5.62 unit cells, in excellent agreement with the reciprocal-space eigenvalue analysis-derived $q_z = 0.178$. This is shown in Figure \ref{fig:bilayer_monolayer_multilayer} b), where we display the mean magnetisation length $\left< m_\mathrm{Ru} \right>$ of one sublattice as a function of the number of unit cells along $c$. The distance between the minima, where spins are fully compensated, reflects the spiral pitch.

\paragraph*{Conclusion.} To summarize, we studied the magnetic groundstate of RuO$_2$ by means of a Wannier function-based mapping from DFT calculations to an isotropic classical Heisenberg model as a function of the Hubbard $+U$ parameter. Our results show that N\'eel temperatures above room temperature are feasible for large choices of $U$, such that $m_\mathrm{Ru} > 0.8\,\mu_\mathrm{B}$. The groundstate of the spin model is helical due to frustration between intra- and inter-sublattice Heisenberg interactions. The approach is validated by comparison with other rutile-structured magnetic materials, for which the correct order and reasonable N\'eel temperatures are determined with the Heisenberg models. Monolayers and bilayers of RuO$_2$ support a collinear altermagnetic order. In the experimental search for signs of altermagnetism in RuO$_2$, it is, according to our results, mandatory to perform measurements at low temperatures: a previous claim of a high N\'eel temperature above 300K together with weak magnetic moments \cite{Berlijn17} is not supported by our model. As a final word of caution, it should be remembered that RuO$_2$ is an itinierant system with "bad" Heisenberg behaviour, i.e. the magnitude of the magnetic moments depends on their mutual alignment. A more involved, self-consistent calculation of reciprocal-space exchange parameters may lead to different conclusions \cite{Daglum2025}.

\subsection*{Appendix: Exchange model, reciprocal-space eigenvalue analysis, and RPA critical temperature}

\paragraph{Exchange model.}
We consider a classical isotropic Heisenberg Hamiltonian written in the form
\begin{equation}
\mathcal{H}
=
-
\sum_{i\alpha,j\beta}
J_{i\alpha,j\beta}\,
\mathbf{\hat{S}}_{i\alpha}\cdot\mathbf{\hat{S}}_{j\beta},
\label{eq:H}
\end{equation}
where $i,j$ label Bravais lattice cells, $\alpha,\beta=1,\dots,n$ label sublattices (basis sites) within the primitive cell, and $\mathbf{\hat{S}}_{i\alpha}$ are classical unit spin vectors. The isotropic exchange parameters $J_{i\alpha,j\beta}$ are obtained from first-principles calculations (TB2J) and depend only on the relative displacement between sites.

Let $\mathbf{R}$ be a Bravais lattice vector and $\boldsymbol{\tau}_\alpha$ the position of sublattice $\alpha$ within the unit cell, such that $\mathbf{r}_{i\alpha}=\mathbf{R}_i+\boldsymbol{\tau}_\alpha$. We define the real-space exchange couplings
$J_{\alpha\beta}(\mathbf{R})\equiv J_{0\alpha,\mathbf{R}\beta}$.
The lattice Fourier transform of the exchange interaction is then given by
\begin{equation}
\mathcal{J}_{\alpha\beta}(\mathbf{q})
=
\sum_{\mathbf{R}}
J_{\alpha\beta}(\mathbf{R})\,
\exp\!\left[
\mathrm{i}\,\mathbf{q}\cdot
\bigl(\mathbf{R}+\boldsymbol{\tau}_\beta-\boldsymbol{\tau}_\alpha\bigr)
\right],
\label{eq:Jq}
\end{equation}
where $\mathbf{q}$ is a wavevector in the first Brillouin zone. For exchange interactions satisfying
$J_{\alpha\beta}(\mathbf{R})=J_{\beta\alpha}(-\mathbf{R})$,
the matrix $\mathcal{J}(\mathbf{q})$ is Hermitian.

\paragraph{Reciprocal-space eigenvalue analysis.}
For each $\mathbf{q}$ on a grid, we diagonalize the $n\times n$ matrix $\mathcal{J}_{\alpha\beta}(\mathbf{q})$ and denote its eigenvalues by $\lambda_\nu(\mathbf{q})$ ($\nu=1,\dots,n$). The largest eigenvalue
\begin{equation}
\lambda_{\max}(\mathbf{q}) \equiv \max_{\nu}\lambda_\nu(\mathbf{q})
\end{equation}
measures the maximal exchange energy gain associated with ordering at wavevector $\mathbf{q}$.
The associated classical energy per unit cell is
\begin{equation}
E_{\mathrm{LT}}(\mathbf{q})
=
-\,\lambda_{\max}(\mathbf{q}),
\end{equation}
and the ordering wavevector $\mathbf{q}_0$ is determined by
\begin{equation}
\mathbf{q}_0 = \arg\max_{\mathbf{q}}\,\lambda_{\max}(\mathbf{q}).
\end{equation}
This procedure identifies the most energetically favorable (possibly incommensurate) magnetic ordering wavevector in frustrated systems.

\paragraph{Random-phase approximation (RPA) critical temperature.}
The same exchange Fourier matrix determines the paramagnetic instability within the Tyablikov (RPA) approximation. Defining the excitation energies relative to the reciprocal-space ordering mode,
\begin{equation}
\Delta_\nu(\mathbf{q})
\equiv
\lambda_{\max}(\mathbf{q}_0) - \lambda_\nu(\mathbf{q}),
\qquad \nu=1,\dots,n,
\label{eq:Delta}
\end{equation}
we determine the multi-sublattice RPA critical temperature by a straightforward extension to the common RPA formula \cite{Sasioglu2005, Sebesta2021}:
\begin{equation}
\bigl(k_{\mathrm{B}} T_c^{\mathrm{RPA}}\bigr)^{-1}
=
\frac{3}{2}\,
\frac{1}{N_{\mathbf{q}}}
\sum_{\mathbf{q}}
\left[
\frac{1}{n}
\sum_{\nu=1}^{n}
\frac{1}{\Delta_\nu(\mathbf{q})}
\right],
\label{eq:RPA}
\end{equation}
where $N_{\mathbf{q}}$ is the number of wavevectors used to sample the Brillouin zone.
Equation~(\ref{eq:RPA}) reduces to the standard single-sublattice RPA expression in the limit $n=1$ and yields a critical temperature that systematically improves upon mean-field theory by accounting for collective spin-wave fluctuations.

For comparison, the mean-field ordering temperature is
\begin{equation}
k_{\mathrm{B}} T_c^{\mathrm{MFA}}
=
\frac{2}{3}\,\lambda_{\max}(\mathbf{q} = 0).
\end{equation}

The analysis is implemented as a \textsc{python}-based command-line tool that is interfaced with the TB2J output files. We make the tool availabe at the github repository TB2J2RPA.

\begin{acknowledgments}
We acknowledge the many open-source software developers and contributors whose efforts make advanced electronic-structure research broadly accessible. Their dedication to creating, maintaining, and improving these tools has been essential for enabling the present work.

\end{acknowledgments}

\section*{AUTHOR DECLARATIONS }

\subsection*{Conflict of Interest}
The authors have no conflicts to disclose.

\section*{Data Availability Statement}

The data that support the findings of this study are available upon reasonable request from the corresponding author.

\section*{References}

\bibliography{cite} 

\end{document}